\documentclass[12pt]{iopart}
\usepackage{graphicx}
\usepackage{epsf,latexsym}

\newcommand{\ket}[1]{\ensuremath{\left| #1 \right\rangle}}

\newcommand{\EX}[1] {\ensuremath{\left\langle #1 \right\rangle}}

\newcommand{\half} {\ensuremath{\frac{1}{2}}}
\newcommand{\ihbar}{\ensuremath{\frac{i}{\hbar}}}
\newcommand{\be}{\begin{equation}}
\newcommand{\ee}{\end{equation}}
\newcommand{\ba}{\begin{eqnarray}}
\newcommand{\ea}{\end{eqnarray}}

\begin{document}

\title[On the correspondence principle in SQUIDs]{On the correspondence principle: implications from a study of the non-linear dynamics of a macroscopic quantum device}
\author{M. J. Everitt}
\ead{m.j.everitt@physics.org}
\address{Centre for Theoretical Physics, The British University in Egypt, El Sherouk City, Postal No. 11837, P.O. Box 43, Egypt.}
\address{Department of Physics, Loughborough University, Loughborough, Leics LE11 3TU, United Kingdom.}

\date{\today}

\begin{abstract}
The recovery of classical non-linear and chaotic dynamics from quantum systems has long been a subject of interest. Furthermore, recent work indicates that quantum chaos may well be significant in quantum information processing.
In this paper we discuss the quantum to classical crossover of a superconducting quantum interference device (SQUID) ring. Such devices comprise of a thick superconducting loop enclosing a Josephson weak link and are currently strong candidates for many applications in quantum technologies. The weak link brings with it a non-linearity such that semi-classical models of this system can exhibit non-linear and chaotic dynamics. For many similar systems an application of the correspondence principle together with the inclusion of environmental degrees of freedom through a quantum trajectories approach can be used to effectively recover classical dynamics. Here we show (i) that the standard expression of the correspondence principle is incompatible with the ring Hamiltonian and we present a more pragmatic and general expression which finds application here and (ii) that practical limitations to circuit parameters of the SQUID ring prevent arbitrarily accurate recovery of classical non-linear dynamics. 
\end{abstract}



\maketitle

\section{Introduction}
\begin{figure}[!t]
\begin{center}
\resizebox*{0.5\textwidth}{!}{\includegraphics{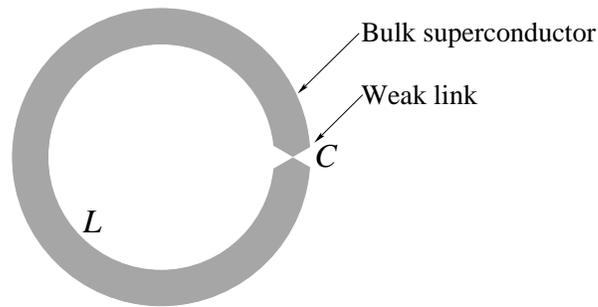}}
\caption{Schematic of a simple SQUID comprising of a superconducting ring (of inductance $L$) enclosing a Josephson junction (with capacitance $C$).
\label{fig:schm}}
\end{center}
\end{figure}

Quantum mechanics, like all physical theories, must satisfy the correspondence principle. That is, in the correct limit every physical theory must reproduce the predictions of classical mechanics. Perhaps because of its many counter-intuitive predictions, demonstrating that the correspondence principle holds for quantum mechanics has long been a subject of interest. One particular class of systems that has undergone extensive investigation comprise those that exhibit non-linear or chaotic dynamics. The interest here arising from the fact that the Schr\"odinger equation is strictly linear. This raises the question of how can non-linear phenomena emerge from such an equation for the evolution of the quantum state? A rather effective answer, that works for both dissipative~\cite{spi1,Spiller95,bru1,bru2,zur1,sch1} and Hamiltonian~\cite{Everitt07} systems is obtained when one introduces the effects of coupling to environmental degrees of freedom and simultaneously invokes the correspondence principle.   
Traditionally, studies of this kind have focused on the measurement problem and foundations of quantum mechanics. However, recent work, that builds on earlier work such as~\cite{Graham:1991p1183,Montangero:2005p1189},  has indicated that understanding such problems will be important in the development of new quantum technologies~\cite{maity06,lages06,kiss06,Rossini06}. Indeed, a recent study has shown that a superconducting quantum interference device (SQUID) based on three Josephson junctions designed for the Josephson flux qubit may be used as a device for the study of quantum chaos~\cite{Dominguez07}.
Such work formed much of our motivation to study the recovery of non-linear trajectories for a related device that is a strong candidate for application in quantum information processing, namely the SQUID ring. 
In this work we consider a SQUID that takes the form of a thick superconducting ring enclosing a Josephson weak link as illustrated in figure~\ref{fig:schm}.

We note that while many textbooks present the correspondence principle as if it was a well understood and closed problem, the reality of the situation is far from this. Indeed, as we have already stated, in this paper we refine a large body of work based on trajectories methods that has seen a great deal of activity over the last fifteen years or so. However, not all systems, even when modelled classically, yield to a trajectory level analysis. For such systems a statistical treatment may be the only possible approach. Hence, expressions of the correspondence principle fall into two categories namely the strong form (trajectory level) and the weak form (statistical and expressed in terms of distributions). While this paper is concerned with the strong form of the correspondence principle we take this opportunity to review an example of recent advances in understanding the quantum classical transition (QCT) in terms of the weak form presented by Greenbaum \emph{et. al.} in~\cite{Greenbaum07}. The authors considered the quantum to classical transition of the chaotic Duffing oscillator. The classical dynamics is described though the evolution of a phase space distribution function that is the solution to the Liouville equation. This is compared with the dynamics of the Wigner function for the corresponding master equation. The authors show that the quantum to classical \emph{``transition occurs due to the dual role of noise in regularising the semiclassical Wigner function and averaging over fine structures in classical phase space''}~\cite{Greenbaum07}. Hence, correspondence was said to be achieved when cross-sectional slices of the Wigner function and the classical distribution function became, to all intents and purposes, indistinguishable. The authors considered the QCT in terms of scaling $\hbar$ -- however we believe that their results are more general and that for non-linear systems the weak form of the QCT is achieved whenever the classical distribution function and the Wigner function become indistinguishable. Although their work is of great interest it does concentrate on the weak form of the correspondence principle. Finally, we observe that Greenbaum \emph{et. al.} note in their work the conditions for weak QCT may need modification for systems with more than one dimension. Our experience modelling coupled quantum systems (especially modelling atom-field collapse and revival phenomena and the QCT to Rabi oscillations~\cite{EverittSM2007}) indicates to us that expressions of the strong correspondence principle of the kind that we present in this work may well be valid for composite systems. However, this is a subject for extensive further study and we will not discuss it further here.

We also take this opportunity to provide a brief overview of some other related work on the quantum to classical transition that we have not previously mentioned. D. Delande on p.665 of~\cite{ChaosAndQM1991} and Friedrich and Wintgen in~\cite{Friedrich:1989p1184} discuss the emergence of chaos in the hydrogen atom. This is done within the context of scaling Planck's constant with respect to the classical action so that \emph{``at small (absolute) values of the energy the quantum spectrum becomes very dense, while it becomes sparse for larger field strengths and binding energies''}~\cite{Friedrich:1989p1184}. Hence, the classical limit is said to be reached when the spectrum is dense and, hence, the effective value of Planck's constant is small. This is somewhat different from our approach where we scale the Hamiltonian (and therefore the spectrum) itself and also impose a localisation condition on the state vector. We also note that the work of M.~Berry and B.~V.~Chirikov that are also in~\cite{ChaosAndQM1991} may well be of interest to the reader.
There are a number of other important works, which mainly focus on solutions to the master equation and weak forms of the QCT, that nevertheless should be mentioned. The QCT for standard map and the quantum ratchet have been studied as dissipative driven systems in~\cite{Dittrich:1990p1188} and \cite{Carlo:2005p1181} respectively. We note that while~\cite{Carlo:2005p1181} is of relevance to weak expressions of the QCT a trajectories approach was used as an aid to computing solutions of the master equation. Finally, in~\cite{Habib:2006p1182} The Duffing oscillator is studied and shown to have a positive Lyapunov even when the system is far from its classical limit. Whilst this study focused on a master equation approach a trajectories technique was used in computing the Lyapunov exponents thus implying the emergence of chaos in terms of a strong form of the correspondence principle. 

SQUID rings, as with other superconducting devices, are unusual in so far as it is quite straightforward to perform experiments on them as a single macroscopic quantum object. Furthermore, they exhibit well known and understood semiclassical behaviour. Hence, as as individual trajectories are accessible, these devices form an ideal testing ground for better understanding the strong forms of the correspondence principle. Hence, this is the focus of this work.

The correspondence principle in quantum mechanics is usually expressed in the form: 
\begin{quotation}
	\emph{``For those quantum systems with a classical analogue, as Planck's constant becomes vanishingly small the expectation values of observables behave like their classical counterparts''}~\cite{Mer98}. 
\end{quotation}
However, and perhaps surprisingly, due to the ring's dependence on Planck's constant we find that this form of the correspondence principle cannot be applied here.  Explicitly, we find that a problem arises when one considers the Josephson coupling term that is flux quantum ($\Phi_0=h/2e$) periodic  in external magnetic flux. The periodicity arises from a topological  quantisation condition reached through an analysis of the  Ginzburg-Landau Cooper pair wave function of the ring. This dependency of the flux quantum on $h$ is highlighted by experiments that leverage this relationship in order to determine Planck's constant (see, for example,~\cite{Frantsuz1992}).  Hence, we arrive at an interesting situation. In order to apply the correspondence principle we would normally wish to shrink $\hbar$ arbitrarily. However, we cannot do this without also changing $\Phi_0$, which is a topological, rather than canonical, quantisation condition. For later reference it is worth presenting a clear statement of the problem: 
\begin{quotation}
There is a semiclassical description for the dynamics of the SQUID ring that is known as the resistively shunted junction (RSJ) model.  It can predict, in some circumstances, results in excellent agreement with experiment. However, despite the Josephson current term containing an explicit dependence on Planck's constant, the model consists of equations of motion in terms of a classical co-ordinate (the flux). Hence, the solution to these equations of motion are classical trajectories. Our problem is then: From a quantum mechanical model of the ring, for a given solution to the RSJ model, how do we reproduce corresponding classical like trajectories in terms of the expectation values of observables?
\end{quotation}

Instead of the usual expression of the correspondence principle we propose a more pragmatic and general statement  which considers that Planck's constant remains fixed and that correspondence must be achieved through varying real circuit parameters. This is expressed as a condition of large dynamics and localisation of the state vector when compared with a Planck cell in phase space. The localisation condition is key for systems whose solutions to the Schr\"odinger equation delocalise over time, such as the SQUID ring, as this requires the introduction of an additional localising effect. We note that unlike many other systems we find that practical limits on circuit parameters prevent us from obtaining an arbitrarily accurate correspondence between the quantum and classical dynamics. A feature that may well be of significance in quantum information processing applications of superconducting circuits.

In this paper we start our discussion by considering the semiclassical resistively shunted junction (RSJ) model of the SQUID ring. This model is traditionally used, with confidence, when the ring capacitance is in the region $10^{-11}$ to $10^{-13}$F~\cite{Likharev86}. We then choose a typical set of circuit parameters and apply a sinusoidal external flux so that strongly non-linear dynamics are obtained. We show the Poincar\'e section of the associated  attractor. In doing this we follow the strategy of much of the previous work done on recovering classical like trajectories of non-linear systems~\cite{spi1,bru1,bru2,zur1,sch1}. We then present the quantum description of this device. Obtaining classical like Poincar\'e section's from the quantum model is achieved following our reformulated version of the correspondence principle. Here, localisation of the wavefunction arises through introducing environmental degrees of freedom via a quantum trajectories method, in this case, quantum state diffusion (QSD). This is an unravelling of the master equation that corresponds to a unit-efficiency heterodyne measurement (or ambi-quadrature  homodyne detection) on the environmental degrees of freedom~\cite{Wis96} (for an excellent introduction see~\cite{Per98,Jacobs06}). It is via such an open systems approach that we obtain both localisation of and non-linear dependence on the state vector from which we expect to be able to recover non-linear dynamics. 
Finally, we show that the quantum jumps unravelling of the master equation can be used to produce equivalent dynamics. This indicates that our results are not restricted to one environmental measurement process and may be quite general in nature. 

\section{The RSJ model}

The SQUID ring is a device that exhibits macroscopic quantum degrees of freedom. However, under certain circumstances, the semiclassical resistively shunted junction (RSJ) model is often applied with a great deal of success. Although the RSJ model contains a term representing the purely quantum phenomena of tunnelling of the supercurrent through the Josephson junction it takes the form of a differential equation for a classical particle. That is, this model describes the evolution of classical trajectories and not state vectors. As with all quantum systems, the recovery of such classical trajectories should arise naturally from the application of the correspondence principle. 

The RSJ equation of motion for the magnetic flux, $\Phi$, within a driven SQUID ring is~\cite{Likharev86}:
\begin{equation}\label{eq:rsj}
C \frac{d^2\Phi}{dt^2}+\frac{1}{R}\frac{d\Phi}{dt}+\frac{\Phi-\Phi_x}{L}+I_c \sin\left(\frac{2\pi \Phi}{\Phi_0}\right)
=I_d \sin \left(\omega_d t \right)
\end{equation}
where $\Phi_x$, $C$, $I_c$, $L$ and $R$ are, respectively, the external flux bias, capacitance and critical current of the weak link, the inductance of the ring and the resistance representing the effects of environmental decoherence. The drive amplitude and frequency are $I_d$ and $\omega_d$ respectively. Finally, $\Phi_0=h/2e$ is the flux quantum. 

We make the following definitions: $\omega_0=1/\sqrt{LC}$, $\tau=\omega_0 t$, $\varphi=(\Phi-\Phi_x)/\Phi_0$, $\varphi_x=\Phi_x/\Phi_0$, $\beta=2\pi L I_c/\Phi_0$, $\omega=\omega_d/\omega_0$, $\varphi_d=I_d L/\Phi_0$ and $\zeta=1/2\omega_0 RC=1/2Q$ where the quality factor $Q=R\sqrt{C/L}$.  We can then rewrite (\ref{eq:rsj}) in the standard, universal oscillator like, form
\begin{equation}\label{eq:rsjNorm}
\frac{d^2\varphi}{d\tau^2}+2\zeta\frac{d\varphi}{d\tau}+\varphi+\frac{\beta}{2\pi} \sin\left[2\pi\left( \varphi+\varphi_x\right)\right]
=\varphi_d \sin \left(\omega \tau \right)
\end{equation}
We note now a property of this model that will become important later when we consider recovering classical like trajectories from quantum mechanical description of SQUID rings. That is, if we make the following transformation
\begin{equation}\label{eq:scale}
C \rightarrow aC \mathrm{\ and\ } L \rightarrow bL
\end{equation}
then we can keep the dynamics invariant by making the following changes at the same time:
\begin{equation}
R\rightarrow\sqrt{{b}/{a}}R, I_d\rightarrow {I_d}/{\sqrt{b}}\ \mathrm{and}\   
\omega_d\rightarrow {\omega_d}/{\sqrt{ab}}. \label{eq:scale2}
\end{equation}

\begin{figure}[!t]
\begin{center}
\resizebox*{0.9\textwidth}{!}{\includegraphics{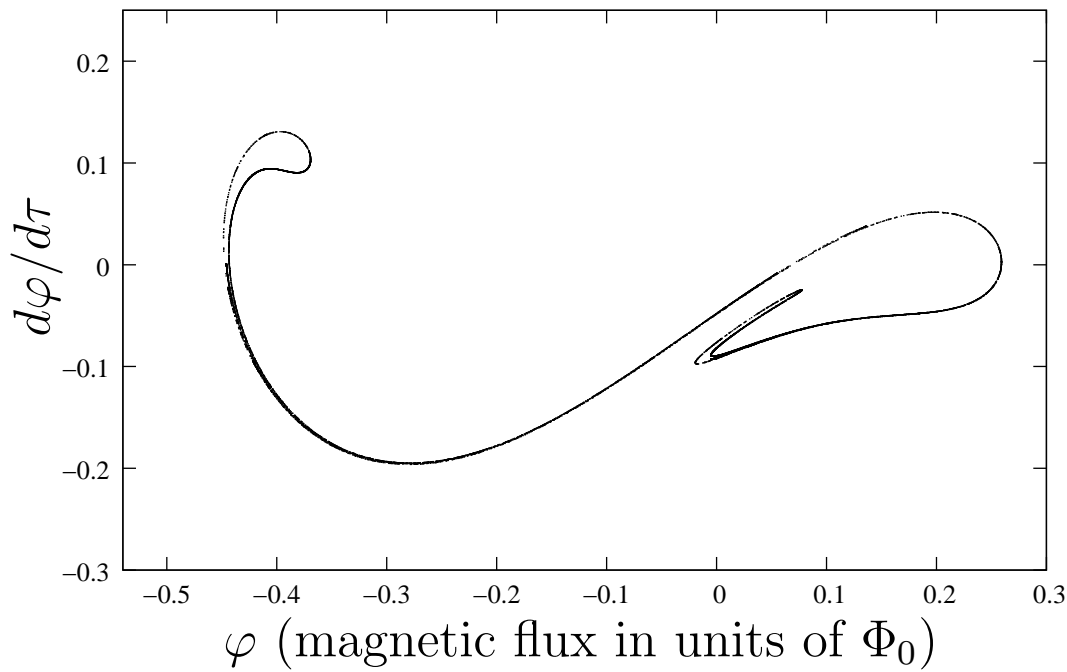}}
\caption{Poincar\'e section obtained from the RSJ dynamics for a SQUID ring with circuit parameters $C=1\times10^{-13}$F, $L=3\times10^{-10}$H, $R=100\Omega$, $\beta=2$, $\omega_d=\omega_0$, $I_d=0.9\,\mu\mathrm{A}$ and $\Phi_x=0.5\Phi_0$.
\label{fig:1}}
\end{center}
\end{figure}

Even though non-linear and chaotic systems can exhibit extremely complex behaviour the Poincar\'e section, comprising of points taken from the phase portrait once a drive period, characterises the dynamics. In figure~\ref{fig:1} we show an example Poincar\'e section for a SQUID ring with circuit parameters $C=1\times10^{-13}$F, $L=3\times10^{-10}$H, $R=100\Omega$, $\beta=2$, $\omega_d=\omega_0$ and $I_d=0.9\,\mu\mathrm{A}$. We note that we have biased the ring at the half flux quantum, $\Phi_x=0.5\Phi_0$, so that the potential approximates a double well.

\section{The quantum model and correspondence}
The quantum description of the SQUID ring is obtained when one considers the following standard Hamiltonian~\cite{Barone1982}:
\begin{equation}\label{eq:qmBase}
\hat{H}=\frac{\hat{Q}^2}{2C}+\frac{\left(\hat{\Phi}-\Phi_x\right)^2}{2L}-
\frac{\hbar I_c}{2e} \cos\left(\frac{2\pi\hat{\Phi}}{\Phi_0} \right)
\end{equation}
which is consistent with the equation of motion~(\ref{eq:rsj}). Substituting $\Phi_0=h/2e$ in order to explicitly show the Hamiltonians dependence on $\hbar$ yields,
\begin{equation}\label{eq:qmBase2}
\hat{H}=\frac{\hat{Q}^2}{2C}+\frac{\left(\hat{\Phi}-\Phi_x\right)^2}{2L}-\frac{\hbar I_c}{2e} \cos\left(\frac{2e \hat{\Phi}}{\hbar} \right)
\end{equation}
where the magnetic flux threading the ring, $\hat{\Phi}$, and the total charge across the weak link $\hat{Q}$ take on the roles of conjugate variables for the system with the imposed commutation relation $\left[\hat{\Phi},\hat{Q}\right]=i\hbar$. For clarity we will use the circumflex to denote operator status  throughout this paper. Here $\Phi_x$ is the external applied magnetic flux and incorporates the drive term for the ring.

We note that from the Hamiltonian one might naively consider that in the limit $\hbar \rightarrow 0$ that the Josephson junction coefficient ${\hbar I_c}/{2e}$ becomes negligible and the system reduces to a harmonic oscillator, where both classical and quantum models could them be made to agree. However, as previously noted, the RSJ model produces a very good match to experiment for situations where ${\hbar I_c}/{2e}$ (and therefore $\hbar$) is non-negligible. It is a  correspondence with such predictions we are interested in recovering.

It is usual, when working with SQUID rings, to normalise flux and charge operators in terms of $\Phi_0$ and $2e$ respectively. However, as $\Phi_0$ is dependent on Planck's constant and $2e$ is not we will not adopt this convention here as doing so obscures the role of $\hbar$. Instead, we define dimensionless flux and charge operators in the manner usual for the simple harmonic oscillator: $\hat{x}=\sqrt{{C\omega_0}/{\hbar}}\hat{\Phi}$ and $\hat{p}=\sqrt{{1}/{\hbar C \omega_0}}\hat{Q}$. These quantities give us a clear framework from which to discuss the correspondence principle. To see why this is the case, let us consider the fact that the commutator $\left[\hat{\Phi},\hat{Q}\right]=i\hbar$ which implies the uncertainty relation $\Delta \hat{\Phi} \Delta \hat{Q} \geq \hbar/2$. This sets a minimum area of uncertainty in phase space, the Planck cell, of $\hbar/2$. Hence, in the conventional view of the correspondence principle, letting $\hbar \rightarrow 0$ is equivalent to making the Planck cell vanishingly small. This implies that as $\hbar \rightarrow 0$ states localised to a Planck cell will look more and more like point particles and effects such as quantum noise fluctuations and tunnelling will become less and less significant.
In dimensionless units the uncertainty relation becomes $\Delta \hat{x} \Delta \hat{p} \geq 1/2$. We will, for future reference, associate a characteristic length with each quadrature of $\delta \hat{x} = \delta \hat{p} = 1/\sqrt{2}$, i.e. the length of one side of a square of area equal to a Planck cell. Hence in this representation the size of the Planck cell is fixed. It is now the system that scales, in this case relative, to $\delta \hat{x}$ and $\delta \hat{p}$.
For example consider the Duffing oscillator, a non linear system with a double well potential. When described in dimensionless units the effect of reducing $\hbar$ is to deepen and broaden the wells in the potential. Hence, increasing the scale of the systems dynamics with respect to the now fixed characteristic length $\delta \hat{x}$. It is this change of scale together with the introduction of an environment that localises the system to a Planck cell that has enabled previous applications of the correspondence principle in terms of $\hbar$ to be so successful~\cite{spi1,Spiller95,bru1,bru2,zur1,sch1}.

Returning to our discussion of the ring Hamiltonian we note that the first two terms are the (shifted) simple harmonic oscillator potential. Hence, normalising $H$ by $\hbar \omega_0$ would imply that the curvature, and therefore spectrum, of such an harmonic oscillator would remain unchanged if $\hbar$ is scaled. Thus, in this system of dimensionless units and normalised energy we can use  $\delta \hat{x}$,  $\delta \hat{p}$ and the background parabolic curvature as a constant reference by which to gauge the effect of changing $\hbar$. This is apparent when we explicitly give the dimensionless Hamiltonian:
$\hat{H}'=\hat{H}/\hbar \omega_0$ we find that
\begin{equation}\label{eq:qmNorm}
\hat{H}'=\frac{\hat{p}^2}{2}+\frac{[\hat{x}-x(t)]^2}{2}-\frac{I_c}{2e \omega_0}\cos \left( \left[\frac{4e^2}{\hbar}\sqrt{\frac{L}{C}} \right]^{1/2} \hat{x}\right)
\end{equation}
Where $x(t)$ represents the normalised external bias flux together with a drive term corresponding to the one used in the RSJ model. We define, for our later convenience, the multiplying factor of $\hat{x}$ within the cosine to be $\Omega=\left[(4e^2/\hbar)\sqrt{(L/C)} \right]^{1/2}$.

It is now that we can see why the SQUID ring does not scale appropriately with $\hbar$. Our discussion of this point makes use of an on-line animation that accompanies this paper. The reader is strongly advised to make use of it as it will greatly clarify the behaviour of the ring as Planck's constant is reduced. 

In the animation we show the potential energy of the ring together with the energy eigenvalues. The eigenvalues are plotted only where the probability density function of the eigenstates are greater than 0.05. Thus we can identify some salient features of the eigenfunctions, such as nodes, without having so much information available that the animation is unclear. Planck's constant is reduced by one percent per frame until it is scaled to approximately $2\times 10^{-6}$ of its original value. We note that this movie uses the dimensionless units described above. Hence the characteristic length  $\delta \hat{x}=1/\sqrt{2}$ and the background parabolic curvature associated with the harmonic oscillator part of the Hamiltonian can be used as a reference in flux and energy respectively. 

At the beginning of the animation we see a double well potential. These wells initially deepen and move closer together with more energy levels being available in each well. About twelve seconds into the animation the wells approach their maximum depth but nevertheless continue to narrow. The result of this is that the number of energy levels in each well begin to reduce. Shortly after this point, as $\Omega$ increases, more wells close in from the outside of the potential. As the animation continues, more and more wells appear in this plot and there are ever fewer energy levels contained in each well (with the zero point energies of each well rising significantly). We also note that there is some tunnelling near the top of the wells. The trend continues until approximately one minute into the animation when the lowest energy eigenstates start to delocalise over a number of wells beyond the central double well. This is due to significant tunnelling under the barrier as the distances between the wells becomes smaller. This trend continues and at the end of the animation we see that the spectrum seems to comprise of two parts - that within the wells and an almost harmonic oscillator type component due to the parabolic background.

Clearly the potential is scaling the wrong way as $\hbar \rightarrow 0$ and we are moving away from a situation where we can recover the type of dynamics seen in Fig.~\ref{fig:1}. Indeed, by looking at the way the spectrum is changing it appears that the system is becoming manifestly more, rather than less, quantum mechanical. Furthermore, if we were to now introduce a localising environment following~\cite{spi1,Spiller95,bru1,bru2,zur1,sch1} then the noise fluctuations, being significant compared with a Planck cell, would simply wash out any details. In this way recovery of a Poincar\'e section as shown in figure~\ref{fig:1} will become impossible. Hence, it is evident that the usual expression of the correspondence principle is incompatible with our objective of recovering RSJ type trajectories for the SQUID ring.

As the notion of correspondence is of fundamental importance we now try to recover the situation and propose a slightly different statement of the correspondence principle from the one already given:
\begin{quotation}
Consider $\hbar$ fixed (it is) and scale the Hamiltonian, in such a way as to preserve the form of the classical phase space, so that when compared with the minimum area $\hbar/2$ in phase space: 
\begin{itemize}
  \item[(a)] the relative motion of the expectation values of any 
             observable (generalised) co-ordinate (and hence the 
             associated classical action) become large and
  \item[(b)] the state vector is localised (in any representation)\footnote{We note that as this item is phrased in terms 
                       of an area in phase space that the localisation 
                       condition must apply to all quadratures. Hence, 
                       in order to aid conceptualisation of this 
                       condition it may be helpful to consider it in 
                       terms of the Wigner function. Here, we would 
                       require this distribution to approximately cover, 
                       and be localised to, the minimum area $\hbar/2$. 
                       Non-negativity of the Wigner function is also 
                       sometimes stipulated as a condition for the QCT. 
                       However, fluctuation terms from interaction with 
                       the environment may generate short lived negative 
                       sections of the Wigner function. Here we 
                       will not impose non-negativity of the Wigner 
                       function as a condition for the QCT as these 
                       interference terms will, in the correspondence 
                       limit, be local and small scale. Therefore, they 
                       will have a negligible effect on the systems 
                       large scale dynamics.}.
\end{itemize}
Then, under these circumstances, expectation values of operators will behave like their classical counterparts.
\end{quotation}
We note that for many Hamiltonians that part (a) here is mathematically equivalent to the original statement of the correspondence principle. That is, as already discussed with reference to the Duffing oscillator, reducing $\hbar$ is the same as scaling the Hamiltonian. However, it is not the only way to scale the Hamiltonian as we can also change the systems parameters.

We have seen that the statement (a) is not the same as reducing $\hbar$ for the SQUID ring. In order to satisfy our new statement of the correspondence principle we see that we must instead scale the Hamiltonian by adjusting circuit parameters. This must be done in such a way so as to preserve the form of the dynamics of the classical phase portrait (i.e. the phase portrait may be scaled relative to a Planck cell but not distorted in any other way). It is now that we see the value of writing the RSJ equation of motion in the dimensionless form~(\ref{eq:rsjNorm}). That is, if we change the capacitance and/or inductance using~(\ref{eq:scale}) and simultaneously make the changes~(\ref{eq:scale2}) that the equation of motion~(\ref{eq:rsjNorm}), and the corresponding phase portrait, remains unchanged. However, we can also see that in terms of the dimensionless flux, $\left\langle \hat{x} \right\rangle$, and charge, $\left\langle \hat{p} \right\rangle$, that the Hamiltonian~\ref{eq:qmNorm} will scale appropriately for the QCT for increasingly large values of capacitance. Hence, we have found a mechanism, via the parameter $a$, for scaling the ring Hamiltonian with respect to the plank cell that avoids the problem we encountered when letting $\hbar$ tend to zero.

We now have an expression of the correspondence principle which encompasses a larger class of systems than under the original expression. Unfortunately, this alone is insufficient to recover classical trajectories. For systems such as the SQUID ring solutions to the Schr\"odinger equation will delocalise to cover much of the possible phase space and expectation values will not behave like their classical counterparts. This can be seen clearly if one looks at phase space representations of the wave function~\cite{Habib98,Greenbaum07}. In order to overcome this difficulty we have introduced the extra constraint (b) of localisation. In this work we localise the wavefunction by treating the SQUID ring as an open quantum system where localisation arises due to the introduction of environmental degrees of freedom.

In an ideal world we would have total control of the parameters of the SQUID ring. Hence, application of the above expression of the correspondence principle could be implemented to arbitrary precision. However, the ring is a real physical device and the above expression of the correspondence principle will be limited by achievable circuit parameters. Typically $C$ takes values between $10^{-11}$F and $10^{-16}$F with the RSJ model being applied for $C=10^{-13}$F and larger. These constraints together with the fact that $\Omega$ scales as $(L/C)^{1/4}$ very much restricts the scalability of this Hamiltonian. That is, there is a practical limit to how classical this device can become.

In any statement of the correspondence principle we expect the expectation values of the observables to behave the same as those of the classical counterpart. By using the Hamiltonian of Eq.~(\ref{eq:qmNorm}) and introducing dissipation as done in the next section we can not produce results that are in agreement with those from the classical equation of motion~(\ref{eq:rsjNorm}). This is due to the fact that this method of introducing Ohmic damping does not bring with it the frequency shift that arises through the damping term in the classical dynamics. We can easily resolve this problem addition of an extra term to the Hamiltonian~\cite{Mer98,Per98}.
That is, the Hamiltonian 
\begin{equation}\label{eq:qmNorm2}
\hat{H}'=\frac{\hat{p}^2}{2}+\frac{[\hat{x}-x(t)]^2}{2}-\frac{I_c}{2 e \omega_0}\cos \left( \Omega \hat{x}\right)+\frac{\zeta}{2}\left(\hat{p}\hat{x}+\hat{x}\hat{p}\right)
\end{equation}
is one we can use to generate trajectories that will be comparable to those predicted by the RSJ model.

\section{QSD and the recovery of classical dynamics}
We now introduce the QSD description for the evolution of the  state vector $\left| \psi\right\rangle$ that is given by the It\^{o} increment equation~\cite{Gis93,Gis93b} 
 \begin{eqnarray}\label{eq:qsd}
 \ket{d\psi}  &  =&-\ihbar \hat{H}' \ket{\psi} dt\nonumber\\
 &&  +\sum_{j}\left[  \EX{\hat{L}_{j}^{\dagger}} \hat{L}_{j}-\half \hat{L}_{j}^{\dagger}\hat{L}_{j}-\half \EX{\hat{L}_{j}^{\dagger}} \Bigl\langle \hat{L}_{j}\Bigr\rangle \right]  \ket{\psi} dt\nonumber\\
 &&  +\sum_{j}\left[  \hat{L}_{j}-\EX{\hat{L}_{j}} \right]  \ket{\psi} d\xi
 \end{eqnarray}
where the first term on the right hand side of this equation provides Schr\"odinger evolution
and the second and third terms describe the decohering effects of the environment on the evolution of the systems state vector.
In this work we consider only one  Lindblad  operator   $\hat{L}=\sqrt{2\zeta}\hat{a}$, where $a$ is the oscillator lowering operator. The time increment is $dt$ and the
$d\xi$ are complex Wiener increments that satisfy $\overline{d\xi^2}=\overline{d\xi}=0$ and       $\overline{d\xi d\xi^{*}}=dt$~\cite{Gis93,Gis93b} (the over-bar denotes the average over stochastic processes). 

\begin{figure}[!t]
\begin{center}
\resizebox*{0.9\textwidth}{!}{\includegraphics{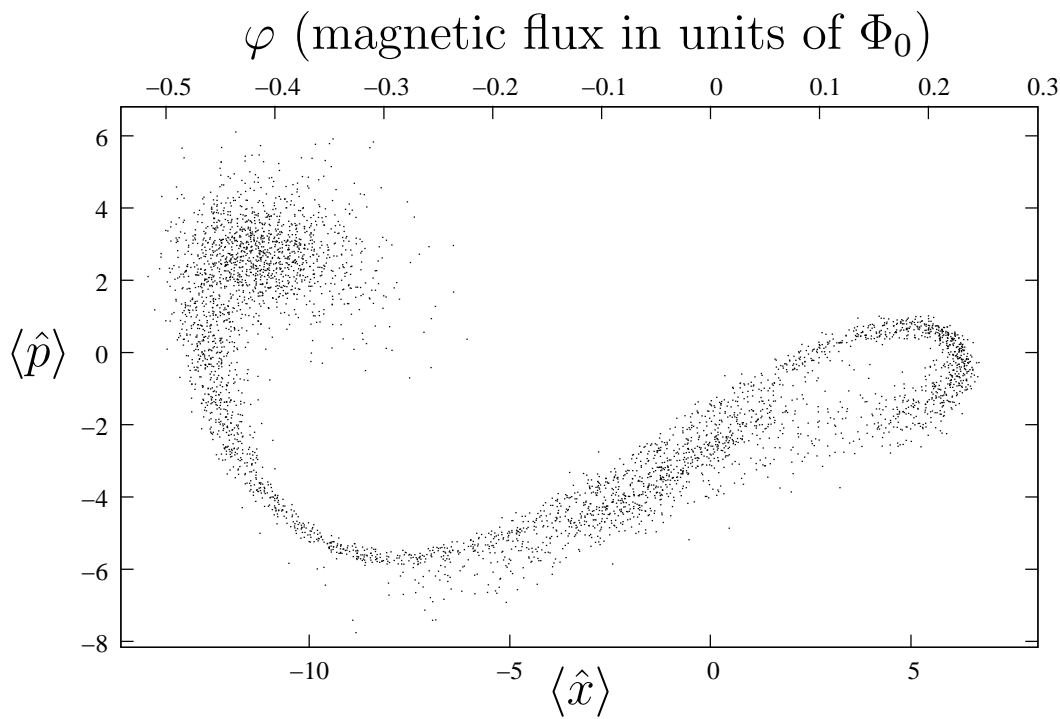}}
\caption{Poincar\'e section obtained from the quantum state diffusion model for the SQUID ring with circuit parameters as given in figure~\ref{fig:1}. As $\left\langle \hat x \right\rangle$ and $\left\langle \hat p \right\rangle$ are dimensionless, we have indicated on top of this figure the magnetic flux in $\Phi_0$ so that direct comparison can be made with figure~\ref{fig:1}.
\label{fig:2}}
\end{center}
\end{figure}

In figure~\ref{fig:2} we show the computed Poincar\'e section obtained for $\left\langle \hat x \right\rangle$ and $\left\langle \hat p \right\rangle$ using this quantum state diffusion model. Here the SQUID ring has the same circuit parameters as given in figure~\ref{fig:1}. We can see that although we have not reproduced exactly the Poincar\'e section of figure~\ref{fig:1} the same structure appears to be indicated here.  

\begin{figure}[!t]
\begin{center}
\resizebox*{0.9\textwidth}{!}{\includegraphics{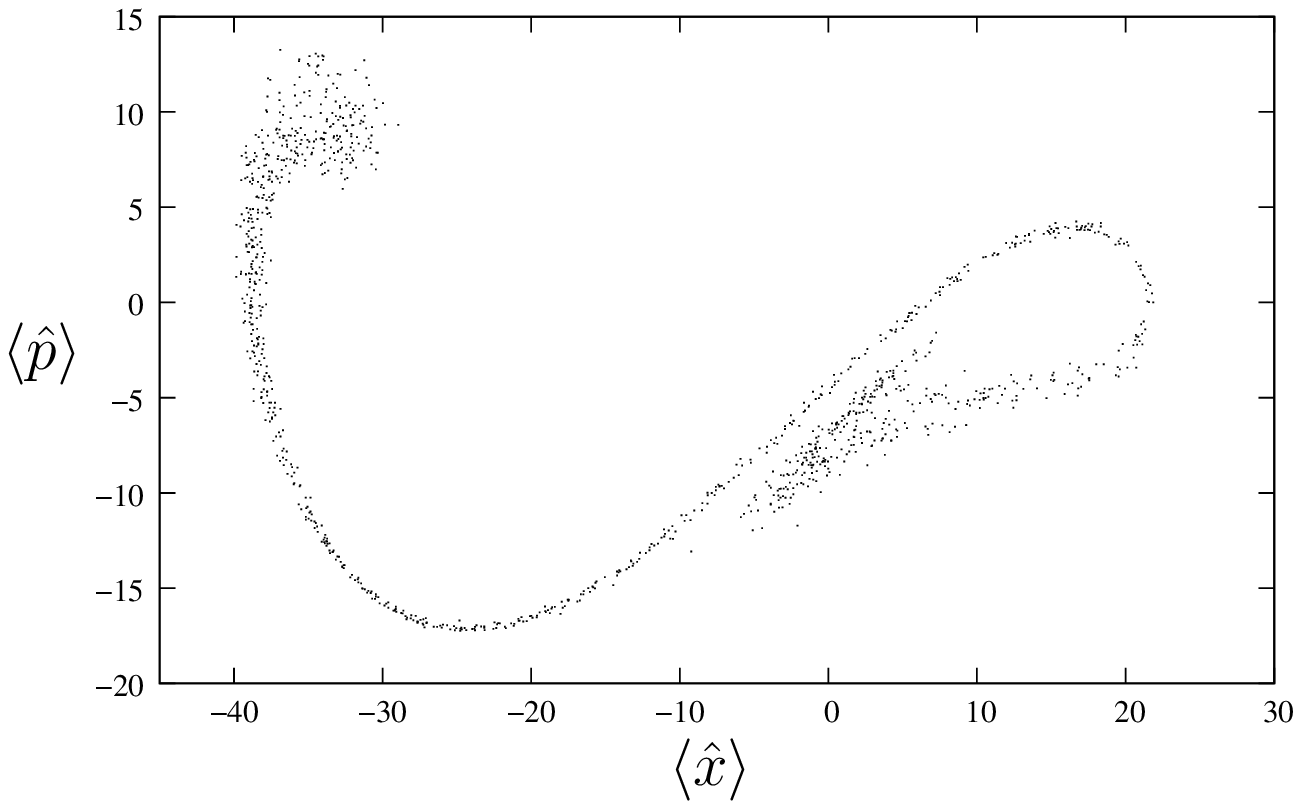}}
\caption{Poincar\'e section obtained from the quantum state diffusion model for the SQUID ring with circuit parameters changed from Fig.~\ref{fig:1} by setting $a=100$ as introduced in equation~(\ref{eq:scale}). This corresponds to increasing the capacitance to $C=10^{-11}$F. 
For clarity we note that this Poincar\'e sections width is in terms of flux quantum approximately the same as that in the section presented in figures 1 and 2. The change in scale, in terms of the dimensionless flux ($\left\langle \hat{x} \right\rangle$) and charge ($\left\langle \hat{p} \right\rangle$), for this plot has arisen through the scaling parameter $a$.
\label{fig:3}}
\end{center}
\end{figure}

From our discussion above we would therefore expect to improve our resolution of this structure for values the scaling parameter $a>1$ where $a$ is defined in equation~(\ref{eq:scale}). We choose $a=100$ so that we obtain a capacitance, $C=10^{-11}$F, commensurate with the the upper practical limit of typical Josephson junctions. In figure~\ref{fig:3} we show the Poincar\'e section for this choice of $a$. It is here that the fact that $\Omega$ scales as $C^{-1/4}$ becomes important. As we can clearly see we have not improved the resolution of detail to the extent that we would have liked. Previous studies of the Duffing oscillator find good correspondence for trajectories~\cite{Bru96,brun97} that cover approximately three times the area of phase space than exhibited in these dynamics. In order to attain this limit we would need a ring capacitance of the order of $10^{-8}$F. Even then some quantum noise would be evident and true convergence to the RSJ model will be achieved only for still greater, and more unrealistic, values of capacitance.

\section{Quantum Jumps}
For  the preceding  results,  we have  used the QSD unravelling of  the master equation. This  unravelling makes assumptions about the underlying measurement process applied by the environment. For QSD this is unit efficiency heterodyne detection  (or ambi-quadrature  homodyne detection)~\cite{Wis96}. Hence, one can only draw limited conclusions from a demonstration of the quantum to classical transition using a specific unravelling. In order to see if our observations of the correspondence limit for the SQUID ring may be general in nature we choose to reproduce the results of figure~\ref{fig:3} using another unravelling of the master equation -- quantum jumps~\cite{Car93,Heg93}. This model is very different from QSD as it is based on a discontinuous photon counting measurement process, rather than a diffusive continuous evolution. In quantum jumps the  pure state stochastic increment equation is given by
\begin{eqnarray}  \label{eq:jumps}
  \left| d \psi \right\rangle & = & - \frac{i}{\hbar}{\hat{H}'}   \left|  \psi \right\rangle dt    \nonumber\\
                                       &    & - \frac{1}{2}\sum_j \left[L_j^\dag L_j -  \left\langle L_j^\dag L_j \right\rangle \right] \left|  \psi \right\rangle dt \nonumber\\
                                       &    & + \sum_j \left[ \frac{L_j}{\sqrt{\left\langle L_j^\dag L_j \right\rangle}}  - 1 \right] \left|  \psi \right\rangle dN_j
\end{eqnarray}
where  $dN_j$   is  a  Poissonian   noise  process  satisfying  $dN_j dN_k=\delta{jk}dN_j$,  $dN_j  dt=0$ and  $\overline{dN_j}=\left\langle  L_j^\dag L_j \right\rangle  dt$. That is jumps occur  randomly at a rate that is determined by $\left\langle L_j^\dag L_j \right\rangle$. Again the  Lindblad  operator is $\hat{L}=\sqrt{2\zeta}\hat{a}$. 
\begin{figure}[t]
\begin{center}
\resizebox*{0.9\textwidth}{!}{\includegraphics{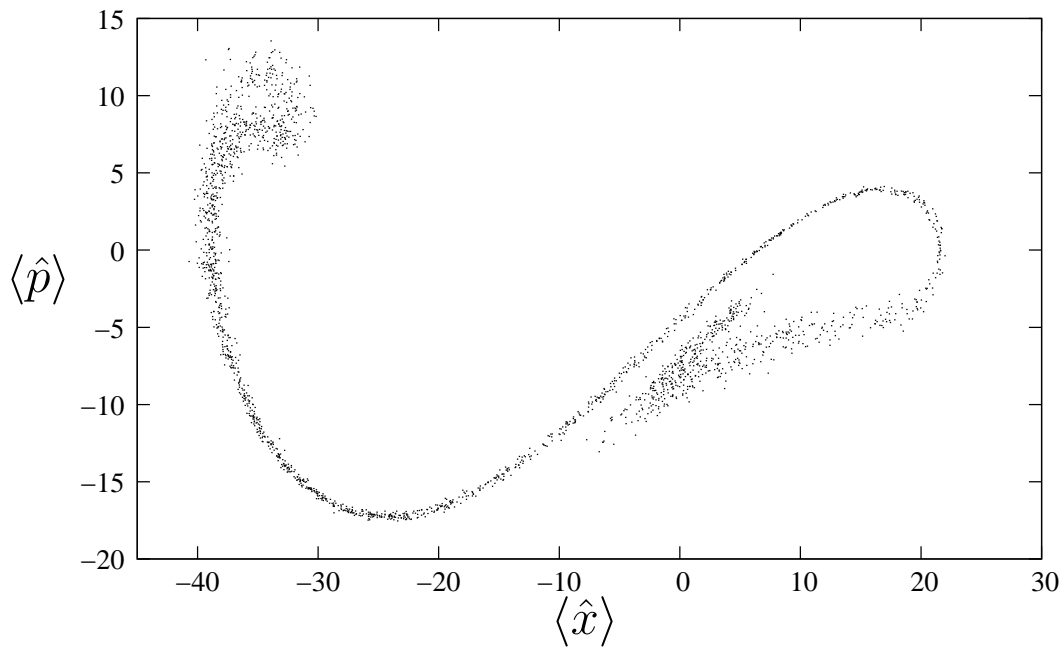}}
\end{center}
\caption{The Poincar\'e section using identical parameters as those used to compute figure~\ref{fig:3} where the evolution is now computed using the quantum jumps unravelling of the master equation instead  of quantum  state diffusion.\label{fig:jumps}}
\end{figure}

If we were to limit our discussion of the QCT of SQUID rings to QSD our results would be meaningful only when the environment operates as a continuous unit efficiency heterodyne detection (or ambi-quadrature  homodyne detection); however, by applying another unravelling such as quantum jumps we can investigate if our results are measurement process dependent. We find that the Poincar\'e sections shown in figures~\ref{fig:3} and~\ref{fig:jumps} generated using QSD and quantum jumps (respectively) are in very good agreement. While it is not possible to model the infinity of possible unravellings of the master equation; the fact that two very different models of environmental measurement processes produce such similar results is, to us, strongly indicative that our results are quite general in nature.

\section{Conclusion}
In this work we have presented a new, pragmatic, reformulation of the strong form of the correspondence principle. The key features being scaling the Hamiltonian (but not through changing $\hbar$) and a localisation of the state vector. Explicitly, in this formulation it is required that the relative motion of the expectation values of observable co-ordinates become large and the state vector is localised when compared with the minimum area $\hbar/2$ in phase space. Then, under these circumstances, expectation values of operators will behave like their classical counterparts. We have used this expression of the correspondence principle to understand the QCT for SQUID rings. Here the classical limit is understood to be correspondence of the dynamics of expectation values of the quantum system with the semi classical RSJ model. To the best of our knowledge this is the first time that the QCT for SQUID's has been demonstrated within a general expression of the correspondence principle. Because we see a QCT for SQUID's it is our opinion that this work reinforces the point of view that the weak or strong formulations of the correspondence principle should not be phrased in terms of scaling $\hbar$.

When our statement of the correspondence principle is applied to the SQUID ring we achieve some success in recovering recognisable classical non-linear trajectories. However, this is somewhat mitigated by practical constraints on circuit parameters. To be precise the Josephson term in the Hamiltonian scales as the fourth root of the ratio of the inductance to the capacitance. It is this dependence that, in practice, prevents arbitrarily good correspondence between the predications of the RSJ model and both the QSD and quantum jumps unravelling of the master equation. Hence we note that quantum effects are therefore non-negligible in the region of parameter space where semi-classical models would normally be used with confidence. One immediate conclusion that can be drawn from this observation is that the the RSJ model will be more effective in situations where any drive amplitude on the ring is large. A good example of such a situation are voltage dynamics experiments where, historically, the RSJ model has been of great utility. Furthermore, this expression of the correspondence principle may well turn out to be of importance when considering Josephson junction devices for application within the field of quantum information processing.

\ack The author would like to thank T.~P.~Spiller, W.~Munro, J.F.~Ralph, T.~D.~Clark, J.~Samson and , S.~Kahlil  for informative and stimulating discussion and  Loughborough University HPC service for use of their facilities.

\bibliographystyle{unsrt}
\bibliography{references}
\end{document}